# Modeling of an EHD corona flow in nitrogen gas using an asymmetric capacitor for propulsion

Alexandre A. Martins
Institute for Plasmas and Nuclear Fusion & Instituto Superior Técnico,
Av. Rovisco Pais, 1049-001 Lisboa, Portugal

Mario J. Pinheiro
Department of Physics and Institute for Plasmas and Nuclear Fusion,
Instituto Superior Técnico, Av. Rovisco Pais, 1049-001 Lisboa, Portugal

## Abstract

The present work intends to identify the nature of the propulsive force that occurs during a positive corona discharge in nitrogen gas using an asymmetric capacitor geometry. We are going to apply the known theory of electrohydrodynamics (EHD) and electrostatics in order to compute all hydrodynamic and electrostatic forces that act on the considered geometry in an attempt to provide a physical insight on the force mechanism that acts on an asymmetrical capacitor.

## Introduction

In this work we investigate the physical origin of the propulsion observed in an asymmetric capacitor that generates an electrohydrodynamic (EHD) flow through a corona discharge in nitrogen gas, at atmospheric pressure. The structure to be studied, an asymmetric capacitor, is represented in figure 1. The notation used for the electrodes is related to their respective radius of curvature. The positive corona wire has a much smaller radius of curvature r than the facing ground electrode *R*.

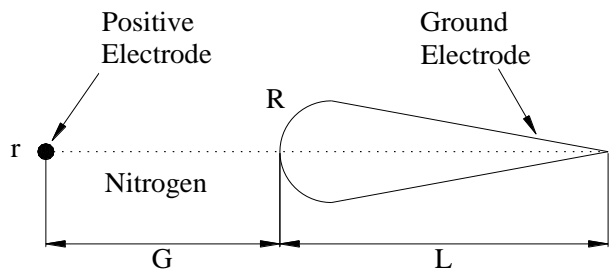

**Figure 1.** Asymmetric capacitor with an air gap G, corona wire r and ground electrode R of length L.

Zhao and Adamiak (2005a, 2006) have studied such geometry with preliminary simulations performed using Fluent and their own code. They attribute the force mechanism mainly to electrostatic forces in the corona wire, directed away from the ground electrode, and state that there are no electrostatic forces on the ground electrode. We disagree with their results since, according to our study, the force on the corona wire has the opposite direction (is directed towards the ground electrode), and the main electrostatic force is precisely on the ground electrode. In the ensuing discussion we are going to clarify these concepts and we are going to apply the known theory of EHD and electrostatics in order to provide a physical insight on the force mechanism that acts on an asymmetrical capacitor.

# Numeric Model

In our numerical model we will consider electrostatic force interactions between the electrodes and the ion space cloud, the moment exchange between the mechanical setup and the induced opposite direction nitrogen flow (EHD flow), nitrogen pressure forces on the structure and viscous drag forces. EHD flow is the flow of neutral particles caused by the drifting of ions in an electric field. In our case these ions are generated by a positive high voltage corona discharge in the high curvature (the higher the radius of a sphere, the less is its curvature) portions of the electrodes. The corona wire has a uniform high curvature and therefore will generate ions uniformly (Chen and Davidson, 2002). On the contrary, the body electrode (ground) has a non uniform curvature having the lowest curvature facing the positive corona wire. The positively ionized gas molecules will travel from the corona wire ion source towards the collector (ground) colliding with neutral molecules in the process. These collisions will impart momentum to the neutral atoms that will move towards the collector as a result. The momentum gained by the neutral gas is exactly equal to the momentum gained by the positive ions accelerated through the electric field between the electrodes, and lost in inelastic collisions to the neutral molecules.

Corona discharges are non-equilibrium plasmas with an extremely low degree of ionization (roughly $10^{-8}$ %). There exist two zones with different properties, the ionization zone and the drift zone. The energy from the electric field is mainly absorbed by electrons in the ionization zone, immediately close to the corona electrode, and transmitted to the neutral gas molecules by inelastic collisions producing electron-positive ion pairs, where the net space charge density $\rho_q$ will remain approximately zero ($\rho_q = 0$). However, the local volume space charge, in the drift zone, will be positive for a positive corona (and therefore $\rho_q = eZ_i n_i$ in the drift zone; $e$ is absolute value of the electron charge, $Z_i$ is the charge of the positive ionic species, $n_i$ is the positive ion density of $N_2^+$) because of the much higher mobility of electrons relative to the positive ions. We will only consider nitrogen ($N_2^+$) positive ions in the drift zone. We make this choice because, for corona discharges in air, the $N_2^+$ and $O_2^+$ ions dominate (Chen and Davidson, 2002), and we want to relate our study with experiments made in an air mixture were nitrogen represents 78% of the total atmospheric gases (22% is Oxygen).

The ionic mobility ($\mu_i$) is defined as the velocity $v$ attained by an ion moving through a gas under unit electric field E (m$^2$ s$^{-1}$ V$^{-1}$), i.e., it is the ratio of the ion drift velocity to the electric field strength:

$$\mu_i = \frac{v}{E}. \tag{1}$$

The mobility is usually a function of the reduced electric field $E/N$ and $T$, where $E$ is the field strength, $N$ is the Loschmidt constant (number of molecules m$^{-3}$ at s.t.p.), and $T$ is the temperature. The unit of $E/N$ is the Townsend (Td), 1 Td = $10^{-21}$ V m$^2$. Since we are applying 28000 V to the corona wire across a gap of 3 cm towards the ground electrode, the reduced electric field will be approximately 38 Td or 38 x $10^{-17}$ Vcm$^2$ (considering that the gas density N at 1 atm, with a gas temperature $T_g$ of 300 K is $N=2{,}447 \times 10^{19}$ cm$^{-3}$). According to Moseley (1969), the mobility $\mu_i$ of an ion is defined by:

$$\mu_i = \mu_{i0}(760/p)(T/273.16), \tag{2}$$

where $\mu_{i0}$ is the reduced mobility, $p$ is the gas pressure in Torr (1 atm = 760 Torr) and $T$ is the gas temperature in Kelvin. For our experimental condition of E/N = 38 Td, Moseley's measurements indicate a $\mu_{i0}$ of 1,83 cm$^2$/(Vs). Thus, at our operating temperature of 300 K, the mobility $\mu_i$ will be 2,01 cm$^2$/(Vs) or 2,01 x 10$^{-4}$ m$^2$/(Vs).

Since the reduced electric field is relatively low, the ion diffusion coefficient $D_i$ can be approximated by the Einstein relation:

$$D_i = \mu_i \left(\frac{k_B T}{e}\right), \tag{3}$$

where $k_B$ is the Boltzmann constant. This equation provides a diffusion coefficient of 5,19 x 10$^{-6}$ m$^2$/s for our conditions.

The governing equations for EHD flow in an electrostatic fluid accelerator (EFA) are already known (Rickard and Dunn, 2007; Zhao and Adamiak, 2005b; Matéo-Vélez et al, 2005) and described next; these will be applied to the drift zone only. The electric field **E** is given by:

$$\mathbf{E} = -\nabla V. \tag{4}$$

Since $\nabla \cdot \mathbf{E} = \frac{\rho_q}{\varepsilon_0}$ (Gauss's law), the electric potential **V** is obtained by solving the Poisson equation:

$$\nabla^2 V = -\frac{\rho_q}{\varepsilon_0} = -\frac{e(Z_i n_i - n_e)}{\varepsilon_0}, \tag{5}$$

where $n_e$ is the negative ion density (we are only considering electrons) and $\varepsilon_0$ is the permittivity of free space. The total volume ionic current density $J_i$ created by the space charge drift is given by:

$$\mathbf{J}_i = \rho_q \mu_i \mathbf{E} + \rho_q \mathbf{u} - D_i \nabla \rho_q, \tag{6}$$

where $\mu_i$ is the mobility of ions in the nitrogen gas subject to an electric field, *u* is the gas (nitrogen neutrals) velocity and $D_i$ is the ion diffusion coefficient. The current density satisfies the charge conservation (continuity) equation:

$$\frac{\partial \rho_q}{\partial t} + \nabla \cdot \mathbf{J}_i = 0. \tag{7}$$

But, since we are studying a DC problem, in steady state conditions we have:

$$\nabla \cdot \mathbf{J}_i = 0. \tag{8}$$

The hydrodynamic mass continuity equation for the nitrogen neutrals is given by:

$$\frac{\partial \rho_f}{\partial t} + \nabla \cdot (\rho_f \mathbf{u}) = 0 \tag{9}$$

If the nitrogen fluid density $\rho_f$ is constant, like in incompressible fluids, then it reduces to:

$$\nabla \cdot \mathbf{u} = 0. \tag{10}$$

In this case, the nitrogen is incompressible and it must satisfy the Navier-Stokes equation:

$$\rho_f \left( \frac{\partial \mathbf{u}}{\partial t} + (\mathbf{u} \cdot \nabla) \mathbf{u} \right) = -\nabla p + \mu \nabla^2 \mathbf{u} + \mathbf{f}. \tag{11}$$

The term on the left is considered to be that of inertia, where the first term in brackets is the unsteady acceleration, the second term is the convective acceleration and $\rho_f$ is the density of the hydrodynamic fluid - nitrogen in our case. On the right, the first term is the pressure gradient, the second is the viscosity ($\mu$) force and the third is ascribed to any other external force $\mathbf{f}$ on the fluid. Since the discharge is DC, the electrical force density on the nitrogen ions that is transferred to the neutral gas is $\mathbf{f}^{EM} = \rho_q \mathbf{E} = -\rho_q \nabla V$. If we insert the current density definition (Equation (6)) into the current continuity (Equation (8)), we obtain the charge transport equation:

$$\nabla \cdot \mathbf{J}_i = \nabla \cdot (\rho_q \mu_i \mathbf{E} + \rho_q \mathbf{u} - D_i \nabla \rho_q) = 0 \tag{12}$$

Since the fluid is incompressible ($\nabla \cdot \mathbf{u} = 0$) this reduces to:

$$\nabla \cdot (\rho_q \mu_i \mathbf{E} - D_i \nabla \rho_q) + \mathbf{u} \nabla \rho_q = 0 \tag{13}$$

In our simulation we will consider all terms present in Equation (13), although it is known that the conduction term (first to the left) is preponderant over the other two (diffusion and convection), since generally the gas velocity is two orders of magnitude smaller than the velocity of ions. Usually, the expression for the current density (Equation (6)) is simplified as:

$$\mathbf{J}_i = \rho_q \mu_i \mathbf{E}, \tag{14}$$

Then, if we insert Equation (14) into Equation (8), expand the divergence and use Equation (4) and Gauss's law we obtain the following (known) equation that describes the evolution of the charge density in the drift zone:

$$\nabla \rho_q \cdot \nabla V - \frac{\rho_q^2}{\varepsilon_0} = 0 \tag{15}$$

In Table I we can see the values of the parameters used for the simulation. We will consider in our model that the ionization region has zero thickness, as suggested by Morrow (1997). The following equations will be applied to the ionization zone only. For the formulation of the proper boundary conditions for the external surface of the space charge density we will use the Kaptsov hypothesis (Kaptsov, 1947) which states that below corona initiation the electric field and ionization radius will increase in direct proportion to the applied voltage, but will be maintained at a constant value after the corona is initiated.

In our case, a positive space charge $\rho_q$ is generated by the corona wire and drifts towards the ground electrode through the gap $G$ (drift zone) between both electrodes and is accelerated by the local electric field. When the radius of the corona wire is much smaller than $G$, then the ionization zone around the corona wire is uniform. In a positive corona, Peek's empirical formula (Peek, 1929; Cobine, 1958; Meroth, 1999; Atten, Adamiak, and Atrazhev, 2002; Zhao, and Adamiak, 2008) in air gives the electric field strength $E_p$ (V/m) at the surface of an ideally smooth cylindrical wire corona electrode of radius $r_c$:

$$E_p = E_0 \cdot \delta \cdot \varepsilon \left(1 + 0.308/\sqrt{\delta \cdot r_c}\right) \qquad (16)$$

Where $E_0 = 3.31 \cdot 10^6 V/m$ is the dielectric breakdown strength of air, $\delta$ is the relative atmospheric density factor, given by $\delta = 298p/T$, where $T$ is the gas temperature in Kelvin and $p$ is the gas pressure in atmospheres ($T$=300K and $p$=1atm in our model); $\varepsilon$ is the dimensionless surface roughness of the electrode ($\varepsilon = 1$ for a smooth surface) and $r_c$ is given in centimeters. At the boundary between the ionization and drifting zones the electric field strength is equal to $E_0$ according to the Kaptsov assumption. This formula (Peek's law) determines the threshold strength of the electric field to start the corona discharge at the corona wire. Surface charge density will then be calculated by specifying the applied electric potential $V$ and assuming the electric field $E_p$ at the surface of the corona wire. The assumption that the electric field strength at the wire is equal to $E_p$ is justified and discussed by Morrow (1997). Although $E_p$ remains constant after corona initiation, the space charge current $J_i$ will increase with the applied potential $V_c$ in order to keep the electric field at the surface of the corona electrode at the same Peek's value, leading to the increase of the surrounding space charge density and respective radial drift.

Atten, Adamiak and Atrazhev (2002), have compared Peek's empirical formula with other methods including the direct Townsend ionization criterion and despite some differences in the electric field, they concluded that the total corona current differs only slightly for small corona currents (below 6 kV). For voltages above 6 kV (corresponding to higher space currents) the difference is smaller than 10% in the worst case, according to them.

For relatively low space charge density in DC coronas, the electric field $E(r)$ in the plasma (ionization zone) has the form (Chen and Davidson, 2002):

$$E(r) = \frac{E_p r_c}{r}, \qquad (17)$$

where $r$ is the radial position from the center of the corona wire. Since the electric field $E_0$ establishes the frontier to the drift zone, using this formula we can calculate the radius of the ionization zone ($r_i$), which gives:

$$r_i = \frac{E_p r_c}{E_0} = r_c \cdot \delta \cdot \varepsilon \left(1 + 0.308/\sqrt{\delta \cdot r_c}\right). \tag{18}$$

Since we have chosen in our simulation for $r_c$ to be 0,025 *mm*, then $r_i$ would be 0,074 *mm*. Now we can calculate the voltage ($V_i$) at the boundary of the ionization zone by integrating the electric field between $r_c$ and $r_i$:

$$V_i = V_c - E_p r_c \ln(E_p/E_0), \tag{19}$$

where $V_c$ is the voltage applied to the corona electrode and $r_c$ is in meters. This equation is valid only for the ionization zone. In our case it determines that if we apply 28000 Volts to the corona wire, then the voltage present at the boundary of the ionization zone becomes 26833,68 Volts.

For the drift zone, Poisson equation (Equation (5)) should be used together with the charge transport equation (Equation (13)) in order to obtain steady state field and charge density distributions. The values of the relevant parameters for the simulation are detailed in Table I.

**Table I. Value of parameters used for the simulation.**

| Parameters | Value |
|---|---|
| Nitrogen density (T=300K, p=1atm), $\rho_N$ | 1.165 kg/m$^3$ |
| Dynamic viscosity of nitrogen (T=300K, p=1atm), $\mu_N$ | 1,775 x 10$^{-5}$ Ns/m$^2$ |
| Nitrogen relative dielectric permittivity, $\varepsilon_r$ | 1 |
| N$_2^+$ mobility coefficient, $\mu_i$ (for E/N = 54 Td) | 1,92 x 10$^{-4}$ m$^2$/(Vs) |
| N$_2^+$ diffusion coefficient, $D_i$ (for E/N = 54 Td) | 4,96 x 10$^{-6}$ m$^2$/s |
| Corona wire radius, $r_c$ | 0,025 mm |
| Facing ground electrode radius, $R_g$ | 1 mm |
| Ground electrode length, L | 40 mm |
| Air gap length, G | 30 mm |
| Corona wire voltage, $V_c$ | 28000 V |
| Ground electrode voltage, $V_g$ | 0 V |

Three application modes of the COMSOL 3.5 Multiphysics software are used. The steady state incompressible Navier-Stokes mode is used to resolve the fluid dynamic equations. The electrostatics mode is used to resolve the electric potential distribution and the electrostatic forces to which the electrodes are subjected. The PDE (coefficient form) mode is used to resolve the charge transport equation (Equation (13)). The parameters used for the simulation are shown in Table I. The structure to be studied is an asymmetric capacitor with one corona wire, centered at (-0.03 m, 0 m), placed at the left of the ground electrode, whose forward section is centered at (0 m, 0 m), as can be seen in figure 2. The solution domain was a square of 0.2 m on each side containing the asymmetric capacitor at the center (figures 2 and 3).

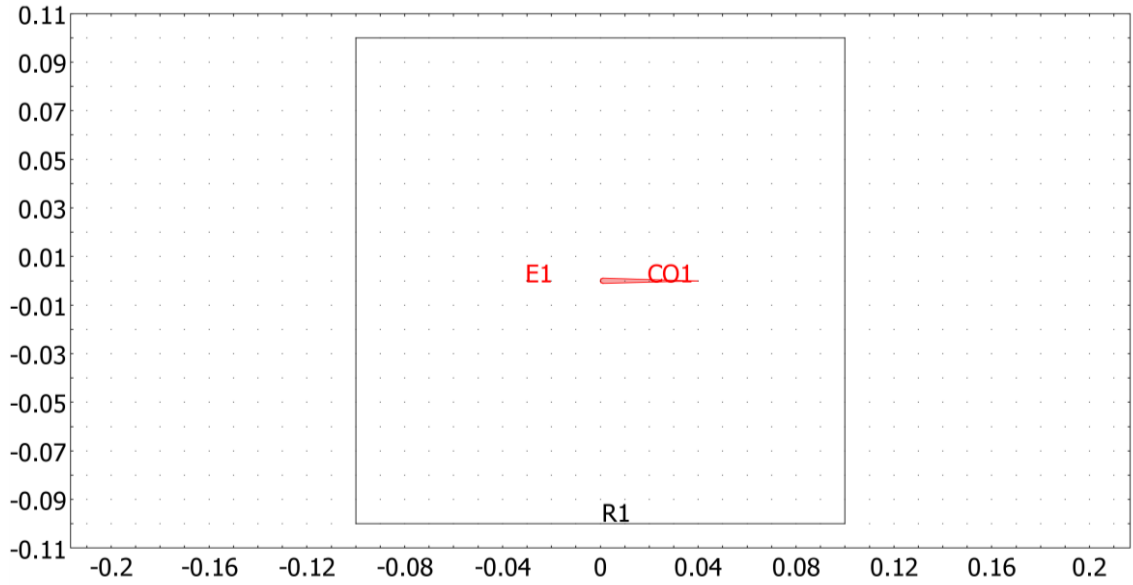

**Figure 2.** Asymmetric capacitor with the corona wire (E1) at (-0.03 m, 0 m) and forward section of ground electrode (CO1) centered at (0 m, 0 m), inside the solution domain (R1) (units in m).

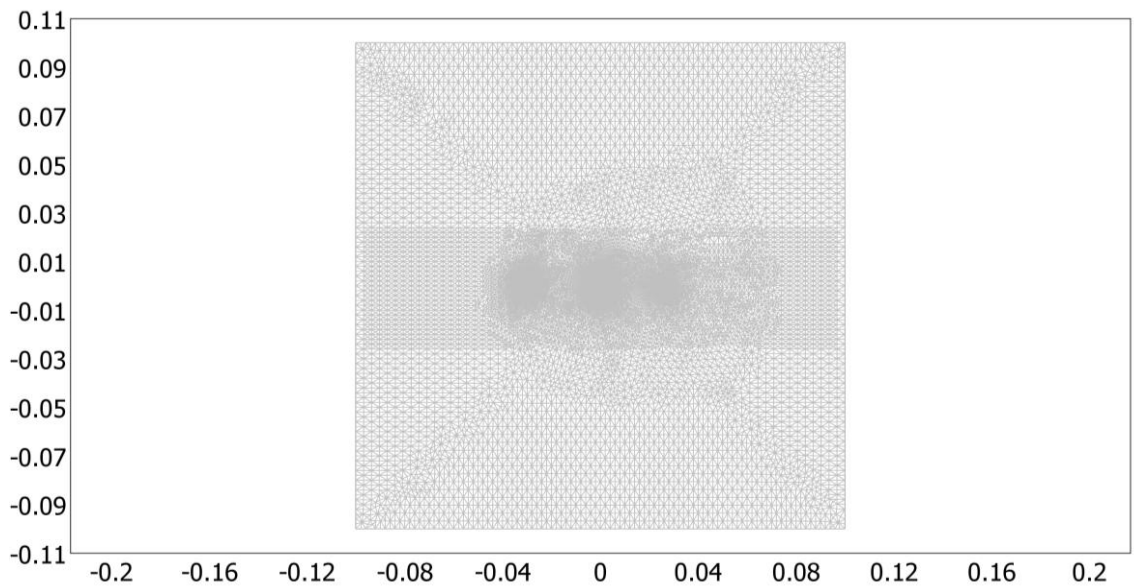

**Figure 3.** Typical mesh of the solution domain (0,2 m×0,2 m) containing 50756 elements (units in m).

## Numerical Simulation Results

Considering the positive space charge distribution around the asymmetric capacitor, we obtain the resulting spatial distribution of the electric field vectors as shown in figure 4.b), which is considerably different from the electric field vectors when there are no positive ions in the surrounding space (figure 4.a)). One can clearly see the shielding effect that the positive ion cloud has on the corona wire electric field and vice-versa. All the forces along the horizontal (x axis) of the lifter are presented in Table II. The results of the simulation show that the electrostatic forces $\mathbf{F_{ex}}$ on the electrodes are the only relevant forces to consider, constituting 99.54% of the total force. The total hydrodynamic force $\mathbf{F_{HTx}}$ (sum of pressure and viscosity forces) is very small because the pressure $\mathbf{F_{px}}$ and viscosity $\mathbf{F_{vx}}$ forces do not contribute in a relevant way in the present conditions. The total resultant force $\mathbf{F_{Tx}}$ (sum of hydrodynamic and electrostatic forces) that acts on the lifter is -0.182 N/m.

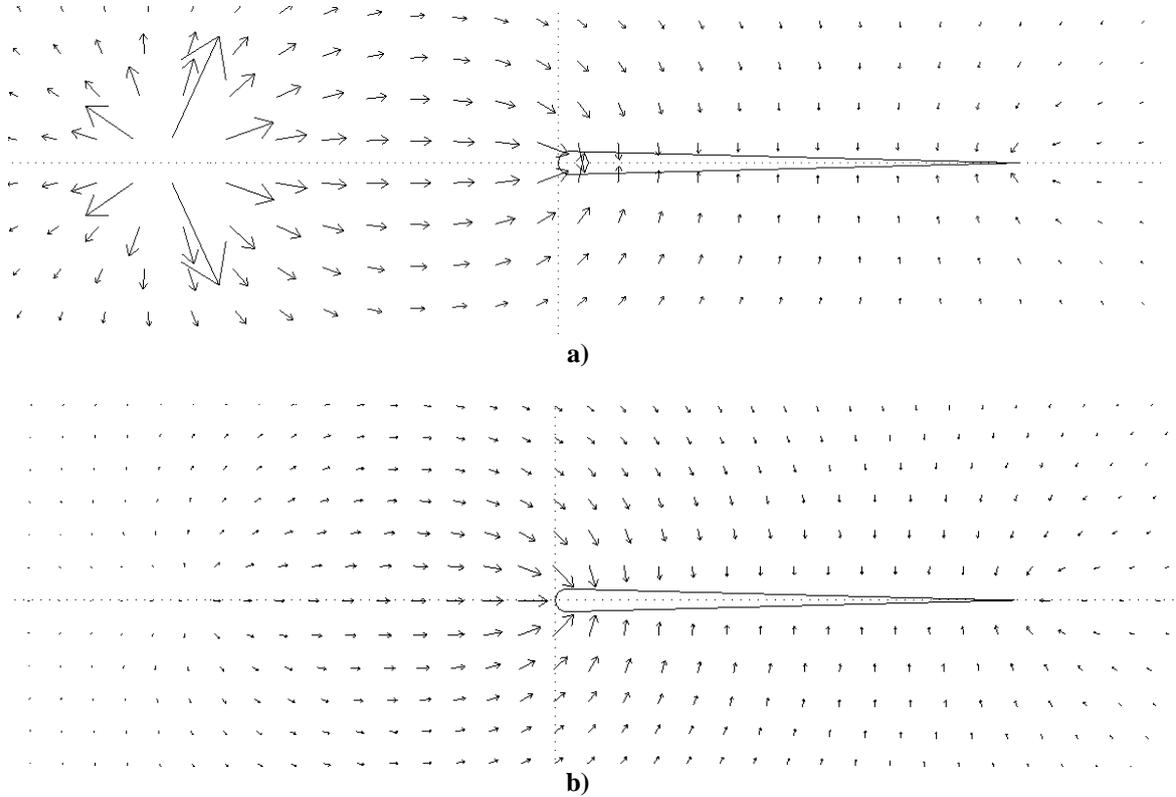

**Figure 4.** Distribution of the electric field vectors (arrows) in the asymmetric capacitor when the corona discharge is not functioning **a)**, and when the corona discharge is functioning **b)**.

**Table II. Forces along the x-axis of the lifter**

|  | $\mathbf{F_{px}}$ (N/m) | $\mathbf{F_{vx}}$ (N/m) | $\mathbf{F_{HTx}}$ (N/m) | $\mathbf{F_{ex}}$ (N/m) | $\mathbf{F_{Tx}}$ (N/m) |
|---|---|---|---|---|---|
| **Corona wire** | -3.642e-5 | -5.142e-5 | -8.783e-5 | 2.279e-4 | 1.400e-4 |
| **Ground electrode** | 8.171e-4 | -1.571e-3 | -7.539e-4 | -1.814e-1 | -1.822e-1 |
| **Total force** | 7.807e-4 | -1.622e-3 | -8.417e-4 | -1.812e-1 | -1.820e-1 |

The corona wire generates a positive charge cloud which accelerates trough the air gap towards the facing ground electrode. The interaction between both electrodes and the positive charge cloud will accelerate the nitrogen positive ions towards the ground electrode and the ions will transmit their momentum to the neutral nitrogen particles by a collision process. Thus, the neutral nitrogen will move in the direction from the corona wire to the ground electrode, and the momentum transmitted to the ions and neutrals is equivalent to the electrostatic forces the ion cloud will induce on both electrodes. The nitrogen neutral wind generated by the collisions with the accelerated ion cloud is shown in figure 5, where the top velocity achieved is 3.426 m/s. The corona wire and nitrogen positive space charge will induce opposite charges in the ground electrode, which will be subjected to a strong electrostatic force. In this way, the real force that makes the asymmetrical capacitor move is not a moment reaction to the induced air draft but moment reaction to the positive ion cloud that causes the air movement. Total momentum is always conserved. The electrostatic force vectors on both electrodes are shown in figure 6.

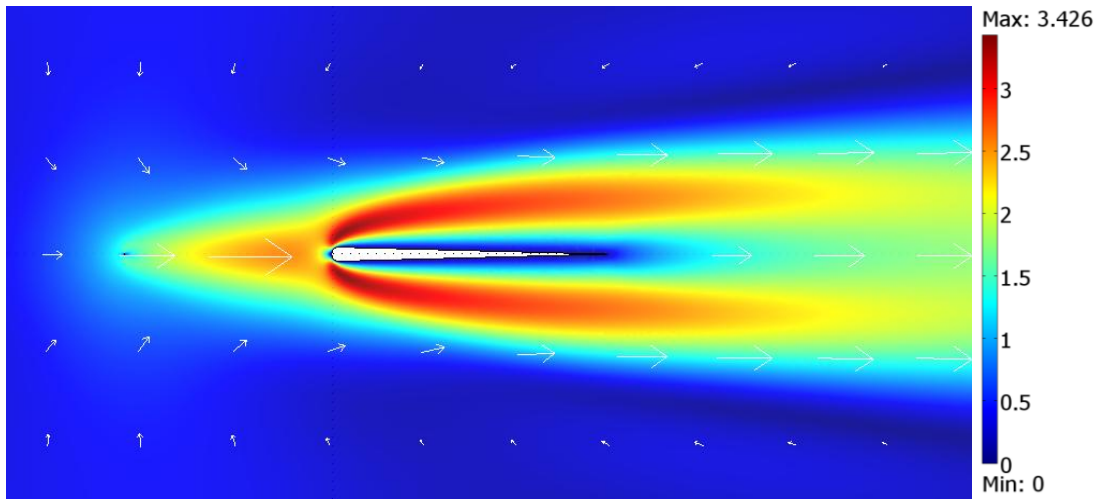

**Figure 5.** Air velocity as surface map with units in m/s, where the arrows are proportional to the magnitude and direction of the air velocity.

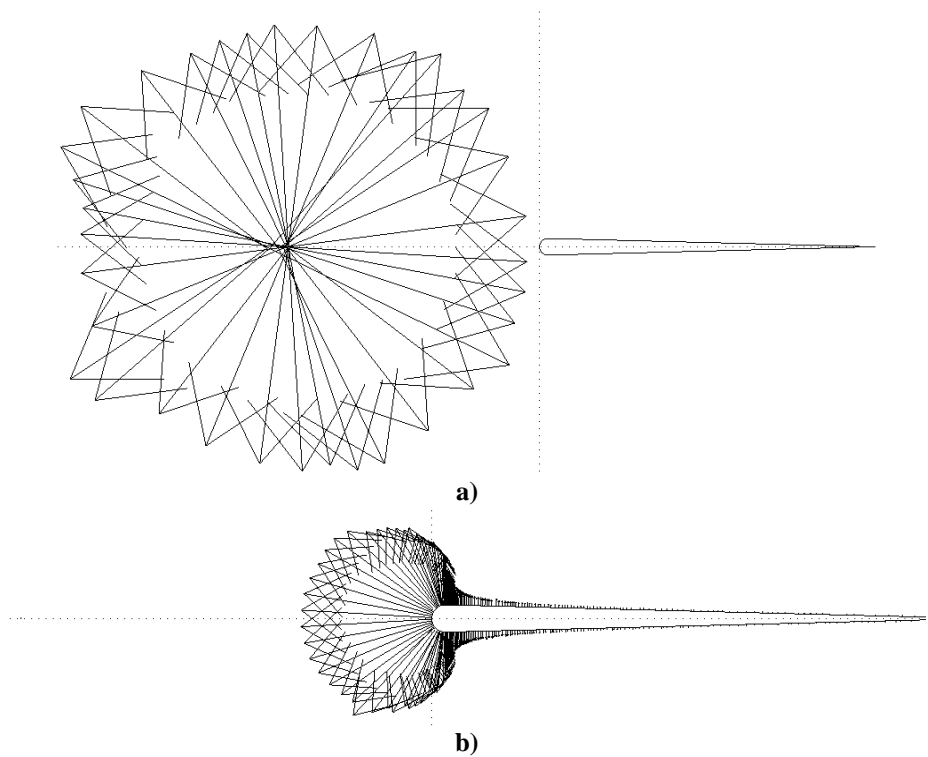

**Figure 6. a)** Electrostatic force on the corona wire, **b)** Electrostatic force on the ground electrode.

The electrostatic force on the corona wire is very strong but mostly symmetric (figure 6.a)). Nevertheless, there is a slight asymmetry in the positive ion cloud distribution around the corona wire (figure 7.a)) in the direction of the ground electrode. The corona wire constantly creates positive ions, which are strongly attracted towards the ground electrode, creating the asymmetrical distribution shown on figure 7.a) and 7.b). Therefore the electrostatic force on the corona wire will be small and directed towards the ground electrode, due to the positive ion distribution around it. On the other hand some positive ions approach and are neutralized on the ground electrode (figure 7.a) and b)), which requires the consumption of a current (electrons) from the power source in order to compensate for the acquired positive charge and remain neutral. As shown by Canning, Melcher, and Winet (2004), in an asymmetrical capacitor of the lifter type, the current from the power source to the ion emitter is always larger than the current to the ground electrode. This is because not all positive ions emitted by the corona wire will be neutralized in the ground electrode; there may be other neutralizing paths. The positive ion cloud around the forward part of the ground electrode (figure 7.a)) induces a charge separation (figure 8) on the ground electrode (which remains neutral in total charge), attracting the mobile electrons towards the places that are surrounded by the positive ion space charge. The electrons that are provided to the neutral electrode (for it to remain neutral) are attracted and neutralized in the front, were they suffer an electrostatic attraction towards the approaching positive ion cloud (figure 7.a)), subjecting the ground electrode to a strong electrostatic force towards the corona wire, or towards the surrounding positive space charge. Therefore, the ion wind is a reaction to the electrostatic thrusting mechanism and not the cause of the thrusting force as it is usually conceived. The main thrust force on the electrodes is electrostatic, not hydrodynamic. The usual mechanical explanation of ion wind momentum is correct, the mechanical momentum relayed to the air by the positive ions is balanced by the mechanical momentum of the physical structure of the capacitor. However, this mechanical explanation does not allow for a perfect understanding of the physical origin of the force that acts on the asymmetric capacitor. The sum of the interaction electrostatic force between the positive space charge ($\mathbf{F_{EI}}$, electrostatic force on the ions) and the capacitor electrodes ($\mathbf{F_{CS}}$, electrostatic force on the capacitor structure) is perfectly balanced:

$$\sum (\mathbf{F_{EI}} + \mathbf{F_{CS}}) = 0. \tag{20}$$

And since the positive ions transmit their momentum to the neutral nitrogen, then we will also have the mechanical momentum conservation:

$$\sum (\mathbf{p_{Nitrogen}} + \mathbf{p_{CS}}) = 0. \tag{21}$$

Where $\mathbf{p_{Nitrogen}}$ is the mechanical momentum that nitrogen acquires from the moving positive ions, and $\mathbf{p_{CS}}$ is the momentum of the capacitor structure. This last equation reflects the generally used ion wind argument, which is correct. However, in order to better understand the physical origin of the force on the capacitor structure we have to recall Eq. (20). The implementation of this equation is not a simple matter. Zhao and Adamiak (2005a, 2006) studied the lifter, and have attributed the observed forces to electrostatic forces on the corona wire, directed away from the ground electrode, but didn´t consider any electrostatic forces on the ground electrode. We have shown that although the ground electrode remains neutral in charge, the surrounding positive space charge induces a charge separation in the electrode (because of the high mobility of electrons in conductors), and this causes a measurable force on the ground electrode. This is proven experimentally by the observed current from the power source to the ground electrode (Canning, Melcher, and Winet, 2004), which has to supply electrons in order to balance the positive ion charge relayed by the surrounding positive ions to the ground electrode, in order to maintain charge neutrality.

This force mechanism remains dependent on the availability of surrounding particles susceptible to ionization in order to be able to function. The measured force in air for a lifter with these dimensions, and a length of 1.2 m was 15,5 g (Chung and Li, 2007). Our calculations for a lifter with the same dimensions in nitrogen provides for a total force of 22,29 g, which is a bit higher (30%) than the force measured in air. From this result, and remembering that nitrogen (considering only $N_2^+$ ions) constitutes 78% of the gases in air, then the force generated in air by the $N_2^+$ ions would be 17,38 g, which is 12% higher than the measured force in air. This represents a fairly good approximation to the force developed in air, considering all approximations made and that besides the $N_2^+$ ions, we also have $N_4^+$, $O_2^+$, $NO^+$ and other ions present in a positive corona in air.

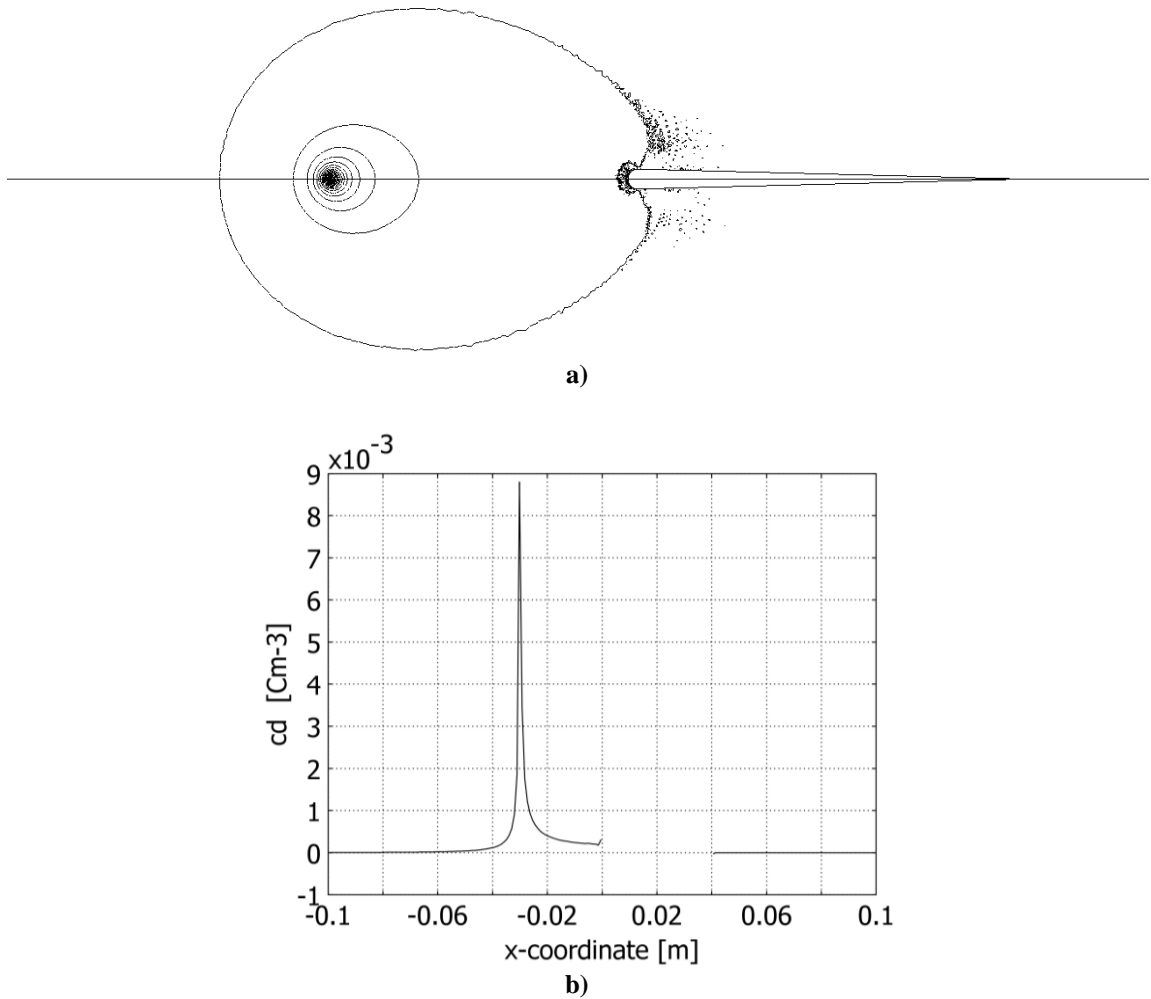

**Figure 7. a)** Lines of equal ionic density, **b)** ionic density distribution cd [Cm$^{-3}$] along the horizontal line (figure 7.a)) through the whole domain from (-0.1 m, 0 m) to (0.1 m, 0 m).

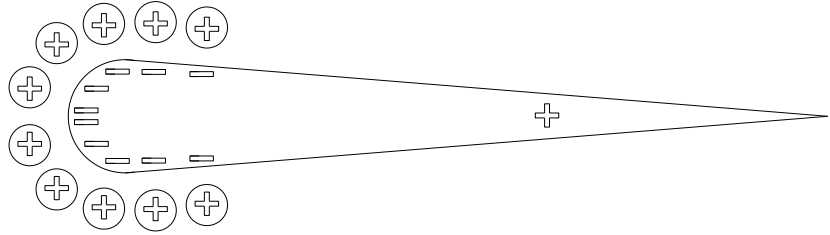

**Figure 8.** Charge separation on the ground electrode by the surrounding positive space charge.

## Conclusion

The main result derived in this work is that the physical origin of the force that acts on an asymmetrical capacitor is electrostatic, and mainly concentrated on the ground electrode (the force on the wire is negligible). Although the usual mechanic explanation of ion wind momentum is correct, it does not allow for a perfect understanding of the physical origin of the force that acts on the asymmetric capacitor. The ion wind is a reaction to the electrostatic thrust mechanism and not the cause of the thrusting force as it is usually conceived. The main thrust force on the electrodes is electrostatic, not hydrodynamic. Nevertheless, this force mechanism is still dependent on the availability of surrounding particles susceptible to ionization in order to be able to function. Even tough our simulation is in nitrogen gas using the $N_2^+$ ion, it provides a reasonable approximation (12% difference) to the same effect in air. The fact that the motive force is directly related to electrostatic forces between both electrodes and the surrounding ion cloud modifies the focus for future development of the effect.

## Acknowledgements

The authors gratefully thank to Mário Lino da Silva for the permission to use his computer with 32 gigabytes of RAM and two quad-core processors, without which this work would not have been possible. We also acknowledge partial financial support by the Reitoria da Universidade Técnica de Lisboa. We would like to thank important financial support to one of the authors, AAM, in the form of a PhD Scholarship from FCT (Fundação para a Ciência e a Tecnologia).